\documentclass[amssymb, preprintnumbers, showpacs, showkeys, aps, prb, superscriptaddress, twocolumn, longbibliography]{revtex4-2}

\usepackage{color} 
\usepackage{xcolor} 
\usepackage{slashed}
\usepackage{graphicx}
\usepackage{amsmath}
\usepackage{latexsym}
\usepackage{epstopdf}
\usepackage{amsmath}
\usepackage{enumitem}
\usepackage{amssymb,amsmath}
\usepackage{multirow}
\usepackage{mathtools}
\usepackage{bbm}
\usepackage{bm}
\usepackage{amsfonts}
\usepackage{amssymb}
\usepackage{calligra}
\usepackage{calrsfs}
\usepackage{makecell}
\usepackage{comment}
\usepackage[normalem]{ulem}
\usepackage[USenglish,american]{babel}
\usepackage{subfigure}

\usepackage[colorlinks,bookmarks=false,citecolor=blue,linkcolor=blue,urlcolor=blue]{hyperref}

\expandafter\ifx\csname package@font\endcsname\relax\else
 \expandafter\expandafter
 \expandafter\usepackage
 \expandafter\expandafter
 \expandafter{\csname package@font\endcsname}%
\fi
\begin{document}
\title{Path Integral Monte Carlo Study of a Doubly-Dipolar Bose Gas}
                
\author{Ratheejit Ghosh}
\email{ratheejit.ghosh@students.iiserpune.ac.in, Corresponding author.}
\affiliation{Department of Physics, Indian Institute of Science Education and Research, Pune 411 008, Maharashtra, India}

\author{Matteo Ciardi}
\email{matteo.ciardi@tuwien.ac.at}
\affiliation{Dipartimento di Fisica e Astronomia, Universit\`a di Firenze, I-50019, Sesto Fiorentino (FI), Italy}%
\affiliation{INFN, Sezione di Firenze, I-50019, Sesto Fiorentino (FI), Italy}
\affiliation{Institute for Theoretical Physics, TU Wien, Wiedner Hauptstraße 8-10/136, 1040 Vienna, Austria}

\author{Rejish Nath}
\email{rejish@iiserpune.ac.in}
\affiliation{Department of Physics, Indian Institute of Science Education and Research, Pune 411 008, Maharashtra, India}

\author{Fabio Cinti}
\email{fabio.cinti@unifi.it}
\affiliation{Dipartimento di Fisica e Astronomia, Universit\`a di Firenze, I-50019, Sesto Fiorentino (FI), Italy}
\affiliation{INFN, Sezione di Firenze, I-50019, Sesto Fiorentino (FI), Italy}
\affiliation{Department of Physics, University of Johannesburg, P.O. Box 524, Auckland Park 2006, South Africa}

\begin{abstract}

By combining first-principles path integral Monte Carlo methods and mean-field techniques, we explore the properties of cylindrically trapped doubly-dipolar Bose gases. We first verify the emergence of a pancake quantum droplet at low temperatures, validating previously mean-field calculations. 
In a regime of small doubly-dipolar interactions, first-principles calculations agree with the generalized Gross-Pitaevskii equation. Such an accordance disappears in a large interaction limit. Here the path integral Monte Carlo estimates the strong doubly-dipolar regime with accuracy. On the contrary, the Gross-Pitaevskii equation does not seize quantum fluctuations in full. We also provide a complete description of the system's quantum behavior in a wide range of parameters. When the system forms a droplet, the superfluid fraction exhibits an anisotropic behavior if compared to the usual Bose gas regime. Interestingly, we observe that the transition temperature from thermal gas to droplet results higher than that of the thermal gas to a Bose-Einstein condensate, indicating the robustness of the droplet against thermal fluctuations. Further, we investigate the anisotropic behavior of the superfluid fraction during the structural transition from a pancake to a cigar-shaped droplet by varying the ratio between electric and magnetic dipole interaction strengths. 
Our findings furnish evidence that the stability of doubly-dipolar Bose-Einstein condensates can be detected in experiments by means of dysprosium atoms.

\end{abstract}


\maketitle

\section{Introduction}\label{intro}

Engineering interatomic potentials via external fields leads to a wealth of exciting phenomena in ultracold quantum gases \cite{Bloch2008, Chin2010}. In particular, systems having anisotropic and long-range dipole-dipole interactions are at the forefront \cite{Defenu2023}. Dipolar quantum gases include magnetic atoms \cite{Chomaz2023, Tang2018}, polar molecules \cite{Moses2017}, and Rydberg atoms \cite{Browaeys2020}. Unlike the magnetic atoms, the latter two systems possess electric dipole moments. A recent study reveals that admixing a pair of low-lying quasi-degenerate states with opposite parity using an electric field can generate an electric dipole moment in a dysprosium atom, otherwise, a magnetic atom. It is a unique scenario in cold atom physics where a single atom simultaneously possesses substantial electric and magnetic dipole moments \cite{lepers2018}. Atoms experiencing a doubly-dipolar potential exhibit highly non-trivial anisotropic properties compared to the usual dipolar interactions from a single dipole moment. Motivated by that, on one side, the role of quantum fluctuations and the formation of droplet states are studied in a doubly-dipolar Bose-Einstein condensate (DDBEC) \cite{chinmayee2020, rathee2022}, and on the other side, quantum spin models are proposed \cite{anich2023}. Remarkably, the doubly-dipolar potential leads to the existence of a pancake droplet, in contrast to the typical cigar droplets so far observed in Bose-Einstein condensates (BEC) experiments \cite{igor2016,chomaz2016, schmitt2016}.

Dipolar interactions critically affect the fundamental properties of a Bose gas \cite{Chomaz2023, Lahaye2009}. Notably, this leads to anisotropic superfluidity \cite{ticknor2011,bismut2012,wenzel2018}, roton-like excitations \cite{chomaz2018,petter2018}, and quantum droplets in dipolar Bose-Einstein condensates (DBECs) \cite{kadau2016,igor2016,schmitt2016}. The latter results from the interplay between contact and dipolar interactions and the stabilization provided by quantum fluctuations. The effect of quantum fluctuations is incorporated in the mean-field approach via the Lee–Huang–Yang (LHY) correction on the chemical potential in the local-density approximation, leading to a generalized Gross-Pitaevskii equation (gGPE). gGPE provides a good qualitative and, to a large extent, quantitative agreement with experiments. However, a better quantitative picture requires going beyond the local-density approximation \cite{Staudinger2018,Fabian2019} or quantum Monte Carlo approaches \cite{9780511614460}. 

Thermal fluctuations also play an essential role in the physics of interacting Bose gases. Exploring finite-temperature effects also requires methods beyond mean-field theory, and path integral Monte Carlo (PIMC) has been successfully employed for such studies \cite{saito2016,PhysRevA.96.013627,fabio2017_2,PhysRevLett.119.215302,piyush2011,nho2005,boninsegni2021}. In addition, PIMC has already proven to be an accurate method to also describe the ground state limit of both weakly and strongly interacting bosonic systems \cite{tobias2020,cinti2019,krauth1996,Pilati2006,holzmann1999,Kora2019}. 
Concerning DBECs systems, PIMC has been instrumental in investigating key macroscopic properties such as the superfluid and condensate fraction \cite{nho2006,piyush2011} (including their critical temperatures \cite{gruter1997,nho2004,ngu2014}), and pair correlation functions \cite{holzmann1999pair}. Additionally, PIMC furnishes a correct description of the supersolid phase in quantum droplets systems \cite{Cinti2010b,Kora2019,cinti14} and Bose glasses too \cite{gau21,ciardi22,PhysRevLett.131.173402}.

In this paper, we analyze the properties of a cylindrically trapped dysprosium doubly-dipolar Bose gas through PIMC simulations based on a continuous-space worm algorithm \cite{boninsegni2006,boninsegni2006_2}. First, we confirm the formation of a pancake droplet at low temperatures, a unique feature first reported in a DDBEC, which also validates the previous results obtained using the gGPE \cite{chinmayee2020, rathee2022}. As expected, the PIMC and gGPE results are in excellent agreement for weak to moderate DDIs. In contrast, for large DDIs, the quantitative deviations are observed, and in particular, PIMC estimates stronger dipolar effects when the $s$-wave scattering length is low. Further, using PIMC, we extract the superfluid fraction as a function of scattering length and temperature. In the droplet regime, we see an enhanced anisotropic superfluid behavior compared to that in the BEC regime because of the highly anisotropic shape of the droplet. Further, we show that the transition temperature for thermal gas to droplet transition is larger than that of the thermal to a repulsive BEC, indicating that the droplet is more robust against thermal fluctuations. Finally, we demonstrate the change in anisotropic behavior in the superfluid fraction under the structural transition from a pancake to a cigar droplet by varying the ratio between the electric and magnetic dipole interaction strengths.

The paper is structured as follows. In sec. \ref{model0}, we introduce the Hamiltonian governing the doubly-dipolar Bose gas and discuss the setup in our system. A comprehensive discussion of the PIMC simulation methodology is provided. Moreover sec. \ref{model0} briefly introduces the gGPE approach too.  We conduct a thorough comparison of the two different methods: gGPE and PIMC simulations in sec. \ref{compare}. Sec. \ref{tr} examines the superfluid behavior of the doubly-dipolar gas with varying short-range interaction.  In sec. \ref{temp}, we focus on the characteristics of the doubly-dipolar condensate and a single pancake droplet under temperature variations. The properties of DDBEC with the varying magnitude of electric dipole moment are discussed in sec. \ref{gamma}. Finally, sec.\ref{con} summarizes the paper's conclusions drawing some future outlooks.

\begin{figure}[t!]
    \centering
    \includegraphics[width=\columnwidth]{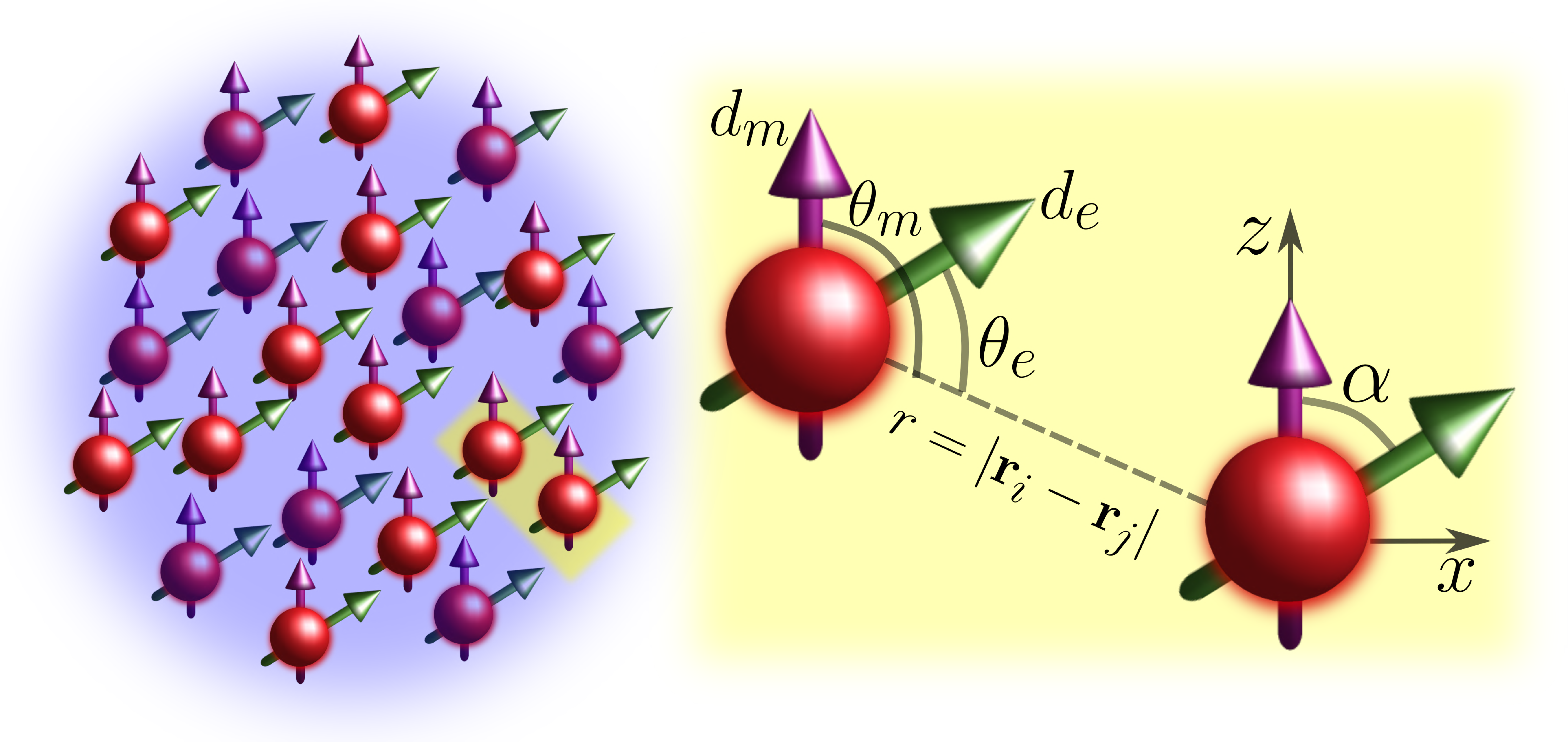}
    \caption{(Color online) Schematic diagram showing Doubly-dipolar Bose gas consisting of $N$ Dy atoms confined in an external harmonic trap, see Section~\ref{model0}. The atoms have both magnetic ($d_m$) and electric dipole moment ($d_e$), the magnetic dipole moment is fixed along the $z$-axis and the electric dipole moment is assumed to be polarized in the $xz$-plane, forming an angle $\alpha$ with the $z$-axis. The angle $\theta_m(\theta_e)$ is the angle between $d_m ( d_e )$  and the vector joining the two atoms, $r=|\textbf {r}_i - \textbf {r}_j|$.} 
    \label{fig:sch}
\end{figure}

\section{Model and Methodology}\label{model0}

We consider a gas of $N$ doubly-dipolar dysprosium (Dy) bosonic atoms of mass $m$ confined in a cylindrical harmonic trap of potential $V(\textbf{r})= m(\omega_x^2 x^2 + \omega_y^2y^2+ \omega_z^2z^2)/2$, where $\omega_q$ is the trap frequency along the $q$th-direction. The atoms possess both magnetic and electric dipole moments polarized by external magnetic and electric fields respectively. The quantum mechanical many-body Hamiltonian describing the system reads
\begin{equation}
\mathcal{H} = \sum_{i=1}^N\Bigg[\dfrac{\textbf{p}_i}{2m} +V(\textbf {r}_i)\Bigg] + \sum_{i<j} U(\textbf {r}_i - \textbf {r}_j) \label{hamiltonian}
\end{equation}
where $\textbf{r}_i=(x_i,y_i,z_i)$ is the position of the $i$-th atom of momentum $\textbf{p}_i$. 
The two-body interaction potential $U$ in Hamiltonian~\eqref{hamiltonian} yields 
\begin{equation}
    U(\textbf {r}_i - \textbf {r}_j) = U_{hard}(r) + U_{DDI}(\textbf {r}_i - \textbf {r}_j)\,. \label{pot}
\end{equation}
where $r=|\textbf {r}_i - \textbf {r}_j|$. $U_{hard}(r)=\infty$ for $r<a_s$ and $U_{hard}(r)=0$ for $r>a_s$, $a_s$ being the two-body s-wave scattering length. $U_{DDI}$ is related to the DDI, which reads
\begin{equation}
    U_{DDI}(\textbf {r}_i - \textbf {r}_j) =g_m\frac{(1-3\cos^2\theta_m)}{{r}^3}+g_e\frac{(1-3\cos^2\theta_e)}{{r}^3}\, , \label{pot}
\end{equation}
where $g_m=\mu_0d_m^2/4\pi$, $g_e=d_e^2/4\pi\epsilon_0$, 
while $d_m$ ($d_e$) is the magnetic (electric) dipole moment of the atoms, moreover  
$\mu_0$ ($\epsilon_0$) is the vacuum permeability (permittivity).
$\theta_m$ and $\theta_e$ are the angles formed by the magnetic and electric dipole vector with the radial vector ${\bf r}$ joining the two dipoles,
and $\alpha$ is the relative angle between the two dipole moments as shown in Figure \ref{fig:sch}. In addition, we define $\gamma= g_e/g_m$ as the ratio that identifies the relative strength between electric and magnetic dipole moments.  

In this work, we assume that the external magnetic field is directed along the $z$-axis, and hence, the direction of the magnetic dipole moment. The external electric field instead varies in the $xz$-plane, inducing a tunable electric dipole moment. The latter forms an angle $\alpha$ with the $z$-axis (see Figure~\ref{fig:sch}). Both $\alpha$ and $\gamma$ can be varied by tuning the external electric field for a range of experimentally feasible values \cite{lepers2018,rathee2022}. Since dipoles are polarized in the $xz$-plane only, the DDI always results repulsive along the $y$-axis and anisotropic in the $xz$-plane.

\subsection{Path Integral Monte Carlo Method}

PIMC is a methodology based on Feynman’s path integral theory \cite{Feymanstat}, which is devised to provide an accurate characterization of the thermodynamic features of quantum systems at finite temperatures. PIMC has been fruitfully applied to a variety of bosonic systems, starting with $^4$He \cite{ceperley1995} and continuing, more recently, with different ultracold atom setups \cite{saito2016,PhysRevA.96.013627,fabio2017_2,PhysRevLett.119.215302,piyush2011,nho2005,boninsegni2021}. In this approach, considering a canonical ensemble, we extract the main information of the bosonic systems by evaluating the partition function $Z$,
which is the trace of the density matrix operator $e^{-\beta \mathcal{H}}$, where $\beta=(k_BT)^{-1}$. 
For $N$ bosons the partition function yields: 
\begin{equation}\label{eq:partf4}
Z = \frac{1}{N!} \sum_{P} \int d\textbf{R}\, \rho(\textbf{R},P\textbf{R},\beta),
\end{equation}
where
\begin{equation}\label{eq:partf2}
\rho(\textbf{R},\textbf{R}',\beta) = \langle \textbf{R} | e^{-\beta H} | \textbf{R}' \rangle
\end{equation}
is the density matrix and $P\textbf{R}$ = $(\textbf{r}_{P(1)},\textbf{r}_{P(2)},\ldots,\textbf{r}_{P(N)})$ is a permutation of particle coordinates $\textbf{r}_i$. 
In this framework bosons at thermal equilibrium are described by their trajectories in imaginary time,  corresponding to bring in a decomposition of Eq.~\eqref{eq:partf2} into a convolution of density matrices at a higher temperature, then Eq.~\eqref{eq:partf4} yielding:
\begin{equation}\label{eq:partf4}
\begin{aligned}
Z=\frac{1}{N!} & \sum_{P}\int  d\textbf{R}^0d\textbf{R}^1\ldots d\textbf{R}^{M-1} \\ &\times\rho(\textbf{R}^0,\textbf{R}^1,\tau) \rho(\textbf{R}^1,\textbf{R}^2,\tau) \cdots\rho(\textbf{R}^{M-1},P\textbf{R}^0,\tau)\,,\\
\end{aligned}
\end{equation}
where $\tau=\beta/M$, while $M$ is the number of time slices, and $\textbf{R}^m$=$(\textbf{r}_{1}^{(m)},\textbf{r}_{2}^{(m)},\ldots,\textbf{r}_{N}^{(m})$ are the coordinates of articles on a given time slice $m$ (with $m=1,2,\dots, M$). By applying this representation, each boson in a position $\textbf{r}_i$ is seen as a classical polymer made of $M$ beads. The entire many-body system can therefore be sampled through the usual Monte Carlo techniques \cite{9780511614460}. Over the years, numerous approximations and sampling schemes have been brought forward, including the so-called worm algorithm to efficiently sample superfluid or condensate fraction \cite{boninsegni2006,boninsegni2006_2}.

PIMC has been applied to three-dimensional dipolar Bose gases before. Nho and Landau \cite{nho2005} first suggested a scheme that employs the Cao-Berne propagator \cite{cao1992} to describe the contact part of the interaction, while adopting a simpler form, the primitive approximation \cite{ceperley1995}, for the long-range component. More recently, Saito \cite{saito2016} has conducted a detailed analysis of this approach, in particular by studying the effect of a short-range cut-off on the $\/r^3$ term, and providing benchmarks \cite{supplemental}. Another possible scheme is to instead describe the contact part of the interaction through a smooth potential, e.g. a Lennard-Jones potential, and treat everything with the primitive approximation \cite{fabio2017_2, boninsegni2021}. The problem of purely repulsive dipoles in two dimensions has also been investigated \cite{buchler2007, filinov2010, piyush2011, tobias2020, cinti2019}.

Generally, dipolar interactions present two numerical complications: first, the long-range character of the interaction leads to a $O(N^2)$ scaling in computational resources; second, the anisotropy of the interaction, as well as its divergent character at $r=0$, require a large number of time slices for convergence in the primitive approximation. In this study, we adopt Saito's scheme \cite{saito2016} adapting it to the DDI. In this way, the density matrix reads
\begin{equation}\label{rhoused}
\begin{aligned}
\rho(\textbf{R}^m, \textbf{R}^{m+1}; & \tau) = \rho_{free} (\textbf{R}^m, \textbf{R}^{m+1}; \tau) \\
&\times\left[\prod_{i<j} \rho_{CB}(\textbf{r}^{(m)}_{ij},\textbf{r}^{(m+1)}_{ij};\tau)\right]\text{e}^{-\tau U_{DDI}(\textbf{R}^m)} \,. 
\end{aligned}
\end{equation}
The first term of the right side of the Eq.~\ref{rhoused} represents the free-particle density matrix:
\begin{equation}
\rho_{free} (\textbf{R}^m, \textbf{R}^{m+1}; \tau)= \frac{1}{(4 \pi \lambda \tau)^{3/2}} \text{e}^{-\tau \frac{(\textbf{R}^m - \textbf{R}^{m+1})^2}{4 \lambda \tau}}\,,
\end{equation}
where $\lambda=\frac{\hbar^2}{2m}$. The second term corresponds to the Cao-Berne propagator \cite{cao1992} 
\begin{equation}
 \rho_{CB} (\textbf{r}, \textbf{r}', \tau) = 1 - \frac{a_s (r+r'- a_s)}{rr'} e^{-\frac{(r-a_s)(r'-a_s)(1+\cos  \widehat{r r^\prime} )}{2\tau\lambda_r}}\,.
\end{equation}
where $\lambda_r=\frac{\hbar^2}{2m}+\frac{\hbar^2}{2m^\prime}$. Finally, we define the exponential argument of the third term as
\begin{equation}
U_{DDI}(\textbf{R}^m) = \sum_{i<j} U_{DDI}(\textbf {r}_i^{(m)} - \textbf {r}_j^{(m)}).
\end{equation}
We run simulations on a cluster, parallelizing the calculation of the interaction to achieve a significant speed-up. This allows us to study large numbers of particles, up to $N=2000$, while ensuring convergence in the discretization of the imaginary-time.

The methodology here introduced is capable of precisely estimating the chief quantities of interacting bosonic systems. In this work, we are interested in calculating the system's energy and its superfluid fraction, both cases
are going to be evaluated at $T>0$ and in the limit of zero temperature. 

Regarding the energy, its estimator relies heavily on the approximation operated on the density matrix. Thus referring to the Eq.~\ref{rhoused}, the total energy reads
\begin{eqnarray} 
\left\langle {\cal H} \right\rangle = \Biggl\langle \frac{3N}{2\tau} & - & \frac{1}{4\lambda\tau^2M}\sum^{M}_{m=1}\left(\textbf{R}^m -  \textbf{R}^{m+1}\right)^2 \nonumber\\
 &-& \frac{1}{M}\sum_{m=1}^{M} \frac{\partial \ln \rho_{pot}(\textbf{R}^{m}, \textbf{R}^{m+1}, \tau)}{\partial \tau}\Biggr\rangle \,, \nonumber \\
 && \label{energytotal}
\end{eqnarray}
$\langle\cdots\rangle$ denoting the statistical mean values of the estimator. 
For the sake of brevity, $\rho_{pot}(\textbf{R}^{m}, \textbf{R}^{m+1}, \tau)$ represents the interacting part of the density matrix, that is second and third terms of Eq.~\eqref{rhoused}.

The superfluid phase can be characterized by the response of the fluid to a small external rotation \cite{Sindzingre1989}. 
While a normal fluid in equilibrium will rotate rigidly with the walls, a superfluid will stay at rest if the walls rotate slowly. The superfluid fraction, $f_s$, can be accurately estimated in a PIMC simulation by calculating the ratio of the moment of inertia of the system to that of the classical moment of inertia\cite{ceperley1995}.

In this work, we evaluate $f_s$ by sampling the well-established ``area estimator''\cite{Sindzingre1989}. This method draws a direct connection between the area enclosed by tangled paths of polymers in a finite system and the reduction of the moment of inertia of the particles compared to the classical case. We inspect the superfluid fraction along three orthogonal axes ($k=\,x,\,y$, and $z$). When doing so, the formula for $f_{s}^{(k)}$ reads:
\begin{equation}
\label{superf1}
f_{s}^{(k)} = \frac{4m^2}{\hbar^2\beta I_{cl}^{(k)}}\left(\langle A_{k}^2 \rangle - \langle A_{k}\rangle^2\right)\,,
 \end{equation}
where we keep the full definition by Sindzingre \textit{et al.} in Refs.~\cite{Sindzingre1989} and \cite{ciardi22}. In this formula, $A_{k}$ results the total area enclosed by particle paths projected onto the plane perpendicular to the $k$-axis, which can be written in terms of particle positions as 
\begin{equation}
\label{superf2}
A_k =  \frac{1}{2} \sum_{i=1}^{N} \sum_{j=0}^{M-1} \left(\textbf{r}_{i}^{j} \times \textbf{r}_{i}^{j+1} \right)_k\,,
\end{equation}
where in this case $\textbf{r}_i^j$ yields the position of the bead corresponding to the $i$-th particle on the $j$-th time slice \cite{zeng2014}. $I_{cl}^{(k)}$ represents the classical moment of inertia around the $k$-axis. It reads
\begin{equation}
\label{superf3}
I_{cl}^{(k)}= m\  \sum_{i=1}^{N} \sum_{m=0}^{M-1} \textbf{r}_{i}^{(j)} \cdot \textbf{r}_{i}^{(j+1)}\,.
\end{equation}
While for an isotropic soft-core fluid, the superfluid fraction along any of the principal directions is the same within errors; the highly anisotropic doubly-dipolar interaction makes $f_s^x$, $f_s^y$ and $f_s^z$ different for a wide range of parameters, depending on the relative angle $\alpha$ and the relative strength $\gamma$.

\subsection{Generalized Gross Pitaevskii Equation}
\label{GPE}
At $T = 0$ the system can be described in the framework of the mean-field theory by gGPE. To incorporate the effect of quantum fluctuation, the LHY correction term is added to the chemical potential, which stabilizes the condensate against mean-filed collapse and helps to describe the beyond mean-field physics \cite{lima2011,lima2012}. The DDBEC including the quantum  fluctuation is described with the generalized GPE: 
\begin{eqnarray}
i\hbar\dot{\psi}(\textbf r,t)=\bigg[-\dfrac{\hbar^2\nabla^2}{2m} +V(\textbf r) + \hspace{2 cm} \nonumber\\
\int d^3r'\psi(\textbf r', t)U(\textbf r-\textbf r')\psi(\textbf r', t)+ \Delta\mu\left[n(\textbf r, t)\right]\bigg]\psi,
\label{nlgpe}
\end{eqnarray}
where $\psi$ denotes the macroscopic condensate wavefunction,  $U(\textbf r)=g\delta(\textbf r)+NU_{DDI}(\textbf r)$ is the interaction potential. The parameter $g=4\pi\hbar^2 a_s N/m$ determines the contact interaction strength, with $a_s$ being the $s$-wave scattering length. The LHY correction term to the chemical potential is given by:
 \begin{equation}\label{lhy}
\Delta\mu=\frac{g_m^{5/2} }{3\pi^3 N}\left(\frac{mn_0}{\hbar^2}\right)^{3/2} \int d\Omega_k\left[\beta+\mathcal F(\theta_k,\phi_k,\alpha)\right]^{\frac{5}{2}}\,,
\end{equation}
where $\int d\Omega_k=\int_0^{2\pi}d\phi_k\int_0^\pi d\theta_k\sin\theta_k$, $\beta = g/N g_m$ and 
\begin{eqnarray}
\mathcal F(\theta_k,\phi_k,\alpha)&=&\frac{4\pi\gamma}{3}\left[3\left(\cos\alpha\cos\theta_k+\sin\alpha\sin\theta_k\cos\phi_k\right)^2-1\right] \nonumber \\
&&+\frac{4\pi}{3}(3\cos^2\theta_k-1)\,,
\label{fd}
\end{eqnarray}
with $\theta_k$ and $\phi_k$ are the angular coordinates in the momentum space. The correction $\Delta\mu$ becomes complex when $\beta<\frac{4\pi}{3}(1+\gamma)$, meaning that the homogeneous DDBEC gets unstable. The real part of $\Delta\mu$ is dominated by hard modes, whereas the unstable low-momentum excitations determine the imaginary part. Not very deep in the instability regime, ${\rm Im}[\Delta\mu]/{\rm Re}[\Delta\mu] \ll 1$ and ${\rm Im}[\Delta\mu]$ can be disregarded.

The LHY correction is repulsive in nature and exhibits a density dependence of $n_0^{3/2}$. Under local density approximation, that is for $n_0\to n(\textbf r, t)$, we can obtain the ground states of a DDBEC by numerically solving Eq.\eqref{nlgpe} via imaginary time evolution. The units of length and energy used for the simulations are $l_y = \sqrt{\hbar/(m \omega_y})$ and $\hbar \omega_y$  respectively.

To test both PIMC and gGPE, Appendix~\ref{densitytest} reports a test for a simple dipolar interaction system.
Moreover, we are going to compare the obtained density profiles with those presented in Ref.~\cite{saito2016}.

\section{BEC-Droplet Transition: PIMC vs gGPE comparison}

\label{compare}
\begin{figure}[t!]
    \centering
    \includegraphics[width=\columnwidth]{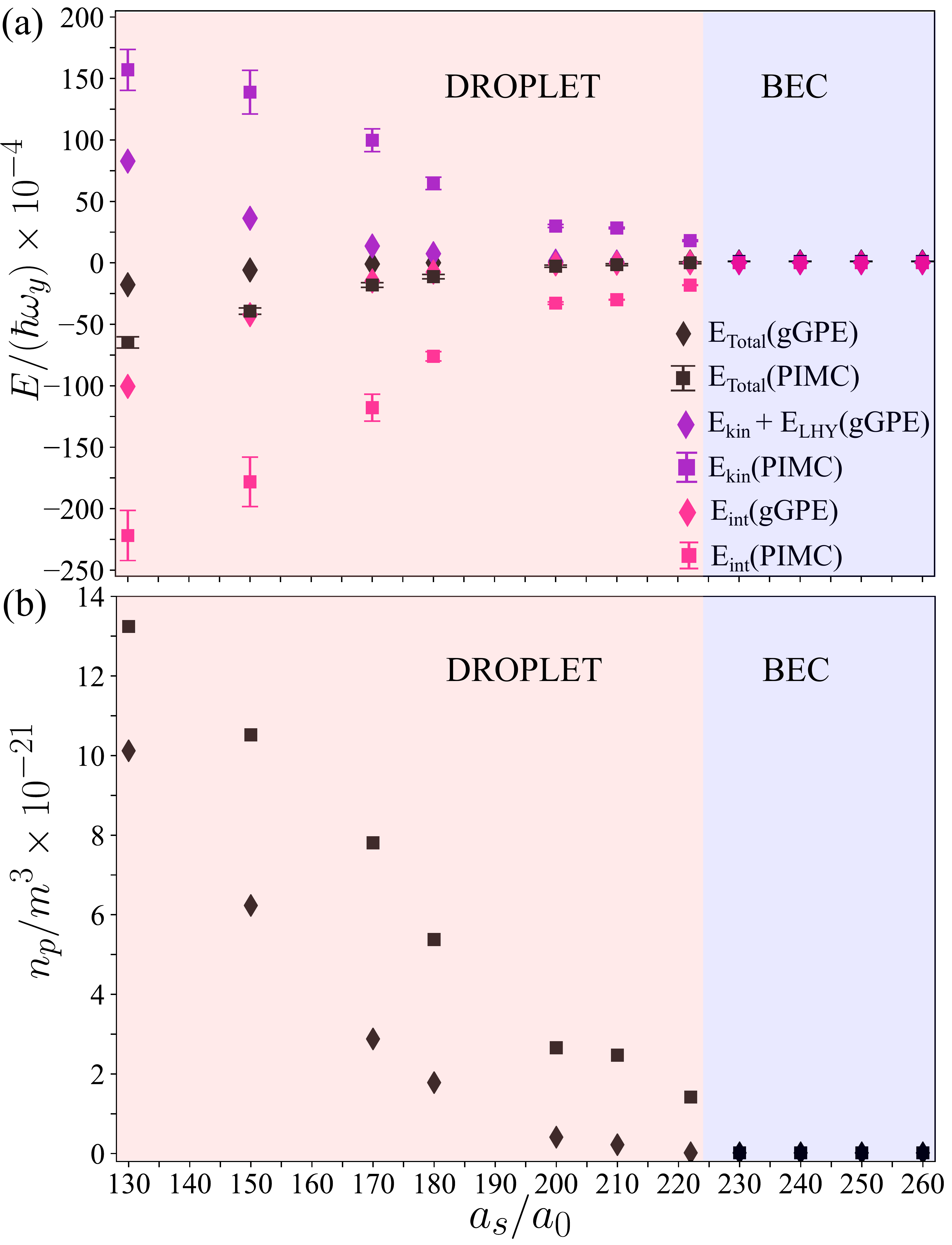}
    \caption{(Color online) (a) Energy comparison from the ground state of gGPE (diamonds) and PIMC (squares). The total energy is shown in black, the energy contributions arising from short-range interactions and long-range DDI are shown in pink, and the energy contributions from the kinetic term and the LHY correction term in gGPE, as well as the kinetic term in PIMC, are illustrated in violet. (b) Peak density comparison from the ground state of gGPE (diamonds) and PIMC (squares).  PIMC simulations have been performed for $N = 1024$ Dy atoms under the external trap, $\omega_{x,y,z}$ = $2\pi\times (75,25,75)$ Hz, see main text. The blue and red background denote the BEC and the droplet regime respectively.}
    \label{figure:2}
\end{figure}

In this section, we analyze the ground state passage of the doubly-dipolar Bose gas from a BEC to a pancake droplet by employing the PIMC method and the gGPE discussed in sec.~\ref{model0}.
It is also interesting to compare the two methodologies under the variation of the short-range interaction and in particular, decreasing the $s$-wave scattering length makes dipolar interactions very dominant.

Regarding PIMC,  we consider a doubly-dipolar Bose gas of $N = 1024$ Dy atoms described by Hamiltonian~\eqref{hamiltonian} and confined in a cylindrical trap elongated along the $y$-axis, with $\omega_{x,y,z}= 2\pi\times (75,25,75)$ Hz. Here, the magnetic and electric dipole moments are assumed to be oriented perpendicular to each other in the $xz$-plane, thus having $\alpha =\pi/2$, and their magnitudes are chosen to result in $\gamma = 1$. Ground state properties are obtained by extrapolating to the limit of zero temperature, that is lowering the temperature until the observables (in this work energy and superfluid fraction) do not change on further decreasing $T$. 
In a different way, gGPE is solved at $T=0$ via imaginary time evolution, see Eq.~\ref{nlgpe}.

We first inspect the ground state energy of the system by varying $a_s$. Figure~\ref{figure:2}a depicts this observable computed by using both methods. Regarding the total energy ($E_{total}$, black points) PIMC and gGPE display a substantially concordant trend between them. For $a_s\gtrsim 222 a_0 $ the repulsive short-range interaction dominates, the system resulting in a stable low-density BEC ground state. As expected, in this region the energy calculations through gGPE exhibit a good agreement with the statistically \textit{exact} PIMC simulations.

On the other hand, for lower scattering lengths the attractive part of the DDI in Eq.~\eqref{pot} dominates over the repulsive short-range one, $U_{hard}$. The \textit{transition} from a BEC to a pancake droplet is observed consistently around $a_s\approx222 a_0$ as witnessed in PIMC. 

The appearance of the droplet regime from a BEC state is also captured well by analyzing the peak density behavior.
A sharp increase in the peak density is evident as the system gets into the droplet regime. The peak density of the droplet rises with a decrease in the scattering length, PIMC and gGPE exhibiting a qualitatively similar pattern as shown in Figure \ref{figure:2}b. The values obtained from PIMC seem to indicate a sharper jump of this quantity around $222a_0$. It is worthwhile to stress that the same tendency can be noted for the kinetic, $E_{kin}$, and potential, $E_{int}$, components of the total energy in Figure \ref{figure:2}a. Differently, gGPE is not able to distinguish the onset of the droplet region through these quantities, showing a smoother behavior. Regarding the PIMC findings, the sharp drop of $E_{int}$ supports the analysis done in Figure~\ref{figure:2}b, indicating the presence of a structure dominated by the DDI. 

The droplet regime features a clear growth of the $E_{kin}$ terms, violet points in Figure~\ref{figure:2}a. 
The LHY correction term combined with kinetic energy in the gGPE simulations results remarkably smaller if compared to PIMC, showing how quantum fluctuations are indeed significant in the regime dominated by strong attractive DDI (low $a_s$). 
As discussed in sec.~\ref{model0}, the introduction of the repulsive LHY correction acts to 
stabilize the condensate against the collapse and form a pancake quantum droplet. However, it should be remembered that LHY only remains a first-order correction to the mean-field theory (one-loop term). It therefore can never account for all the terms that characterize quantum fluctuations as PIMC can properly. This explains the remarkable mismatch of $E_{kin}$ between the mean-filed calculations and the PIMC simulations in the droplet region.

\begin{figure}[t!]
    \centering
    \includegraphics[width=\columnwidth]{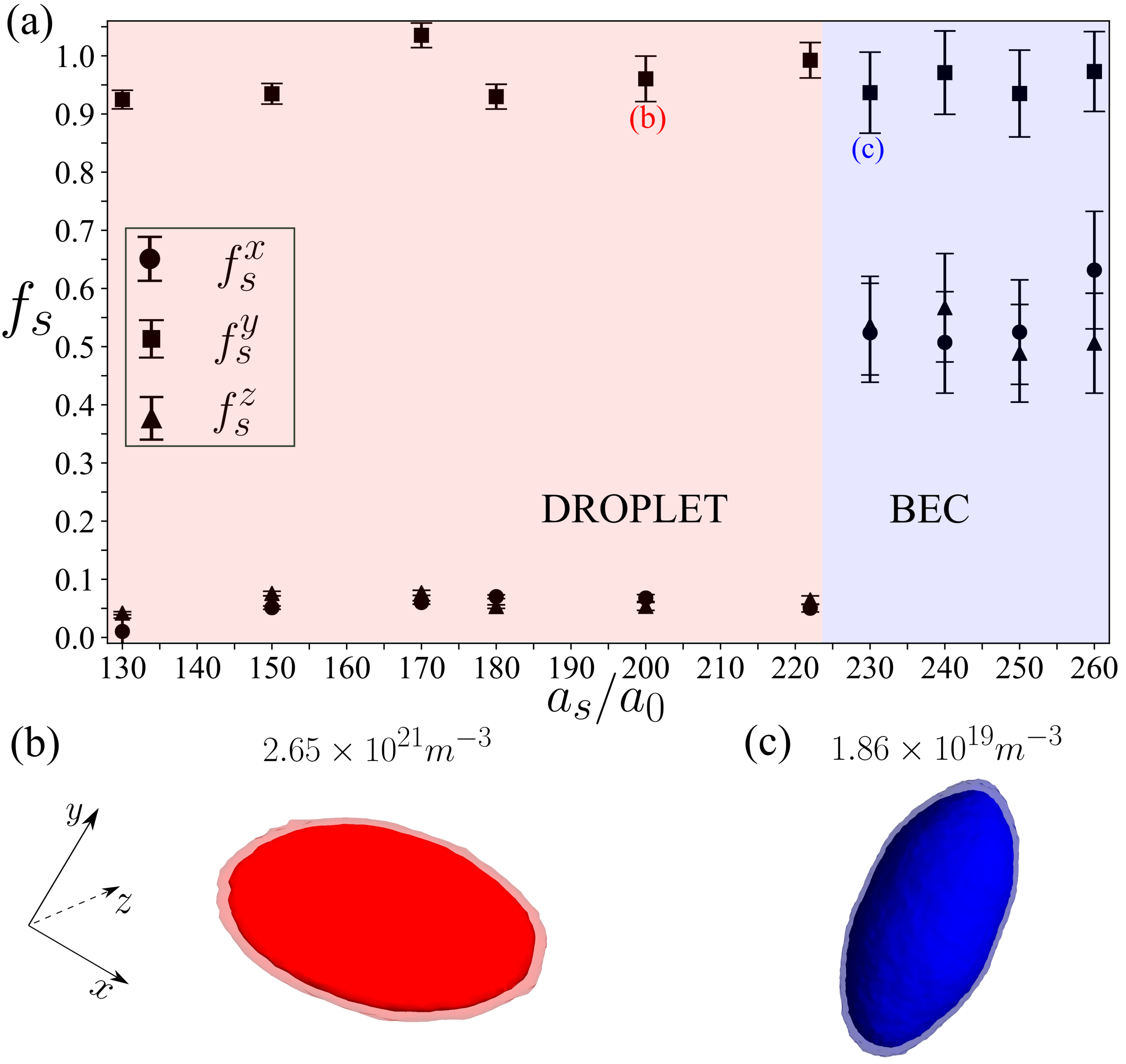}
    \caption{(Color online) (a) Superfluid fraction across BEC to droplet transition along three orthogonal directions $(x,y,z)$ with varying scattering length $a_s$ for $\alpha = \pi/2$, and $\gamma$ = 1, for a trap frequency, $\omega_{x,y,z}= 2\pi\times (75,25,75)$\,Hz. The blue and red background denote the BEC and the droplet regime respectively. We show the 3d density isosurfaces obtained via PIMC simulations, for (b) a pancake droplet ($a_s/a_0$= 200), (c) a cigar-shaped BEC ($a_s/a_0$= 260). The peak densities for each case are provided at the top. The PIMC simulations are done for $N = 1024$ number of Dy atoms. }
    \label{figure:3}
\end{figure}

\begin{figure}[t!]
    \centering
    \includegraphics[width=\columnwidth]{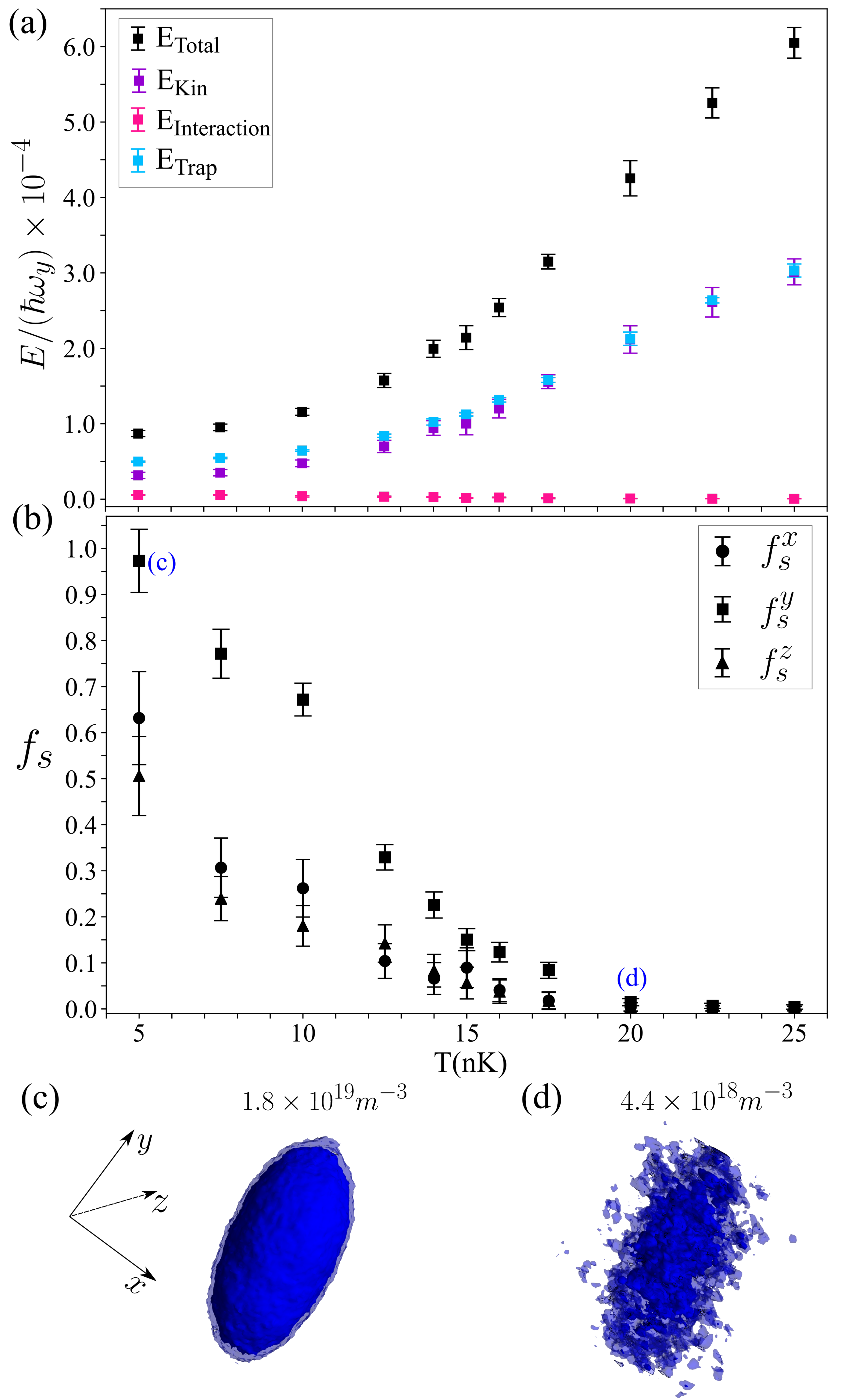}
    \caption{(Color online) (a) Energy variation with temperature for a doubly-dipolar condensate at $a_s/a_0 = 260 $. The total energy is shown in black, the energy contributions arising from short-range interactions and long-range DDI are shown in pink, and the energy contributions from the kinetic term and external harmonic trap are illustrated in violet and skyblue respectively.  (b) Superfluid fraction along the transition from a thermal gas to a doubly-dipolar condensate. The 3d isodensity surfaces are shown for (c) a cigar-shaped doubly-dipolar condensate (T = 5 nK) and (d) a thermal gas (T = 20 nK). The peak densities for each case are provided at the top. The PIMC simulations are done for N = 1024 Dy atoms for an external harmonic trap, $\omega_{x,y,z}$ = $2\pi\times (75,25,75)$ Hz. } 
    \label{figure:4}
\end{figure}

\section{Superfluid Fraction across BEC-Droplet transition }\label{tr}

One important macroscopic observable that characterizes the quantum properties of the doubly-dipolar Bose gas is the superfluidity. We compute the superfluid fraction across the BEC-to-droplet transition via PIMC simulations employing area estimator methods as defined in Eqs.~\eqref{superf1}, \eqref{superf2}, and \eqref{superf3}. The upper panel of Figure \ref{figure:3} illustrates the ground state limit superfluid fraction along the three orthogonal directions ($x,y,z$) with varying short-range interactions. The density distributions obtained from PIMC simulations are visually represented through isodensity plots in the lower panel of Figure \ref{figure:3}.

For higher values of $a_s$, the condensate represents the ground state of the system, which has a three-dimensional cigar shape elongated along the $y$-axis (see Figure \ref{figure:3}c), determined by the underlying trap geometry. The superfluid fraction values computed in the condensate regime at $T= 5$\,nK, are finite in all three directions, with $f_s^{y}$ reaching unity and $f_s^{x}$ and $f_s^{z}$ having lower values. The anisotropy in the superfluidity is due to the anisotropic external harmonic trap applied to confine the doubly-dipolar gas. The high values of the superfluid fraction indicate the inherent superfluid nature and the quantum coherence present in the DDBEC.

With decreasing $a_s$, the superfluid fraction shows a drastic change and becomes highly anisotropic. In this regime, the interaction is dominated by the strong DDI, which is purely attractive in the dipole plane ($xz$) and repulsive perpendicular to the dipole plane ($y$-axis). The attractive DDI forms a quasi-two-dimensional pancake-shaped droplet extending in the $xz$-plane (see Fig.\ref{figure:3}b), which emerges as the ground state of the system. Along the direction perpendicular to the dipole plane, the superfluid fraction reaches unity, while its values are greatly suppressed within the dipole plane. At a temperature of $T = 10$\,nK, $f_s^{(y)}$ achieves around $100\%$ superfluidity, and we extrapolate these values to lower temperatures.  The anisotropy of the superfluidity in this regime is entirely dictated by the anisotropic DDI, with minimal influence from external trap geometry. Similar to the DDBEC, the superfluid fraction emphasizes the intrinsic superfluid nature of the doubly-dipolar pancake droplet.

\section{DDBEC with varying Temperatures}
\label{temp}
In this section, we study the properties of the doubly-dipolar condensate and the pancake droplet with temperature variations. 

\subsection{A Doubly-Dipolar Condensate}

As discussed in previous sections, at high scattering lengths, the doubly-dipolar gas resides in a weak DDI regime. We compute the energy and the superfluid fraction across the passage from a thermal gas to a repulsive doubly-dipolar condensate.
   
At high temperatures, the thermal fluctuations dominate and the system describes a simple thermal gas, without displaying any coherence between the particles. The pronounced thermal fluctuations result in a high value of the kinetic energy as shown in Figure \ref{figure:4}a. The detailed energy analysis shows that the energy due to the harmonic trap is much higher compared to the interaction energy resulting in a cigar-shaped geometry of the thermal gas dictated by the cylindrical trap (see Figure \ref{figure:4}d). As the temperature diminishes, the thermal fluctuations decrease, and below a critical temperature, the thermal gas undergoes a transition to a condensed state. The transition is captured through a sharp increase in the superfluid fraction value around $T = 18$\,nK (See Figure \ref{figure:4}b). Superfluidity increases in all three orthogonal directions as temperature decreases and reaches $100 \%$ superfluidity along the $y$-direction at lower temperatures. In this repulsive BEC regime, the role of doubly-dipolar interaction is minimal, and the anisotropy of superfluidity is attributed to the anisotropic trap geometry. At lower temperatures where thermal fluctuations are negligible, the system converges toward its ground state, with the total energy exhibiting minimal variation.  Figure \ref{figure:4}c illustrates an isodensity plot of a superfluid cigar-shaped condensate at $T = 5$\,nK. 
         
\subsection{A Doubly-Dipolar Pancake Droplet}

\begin{figure}[t!]
    \centering
    \includegraphics[width=\columnwidth]{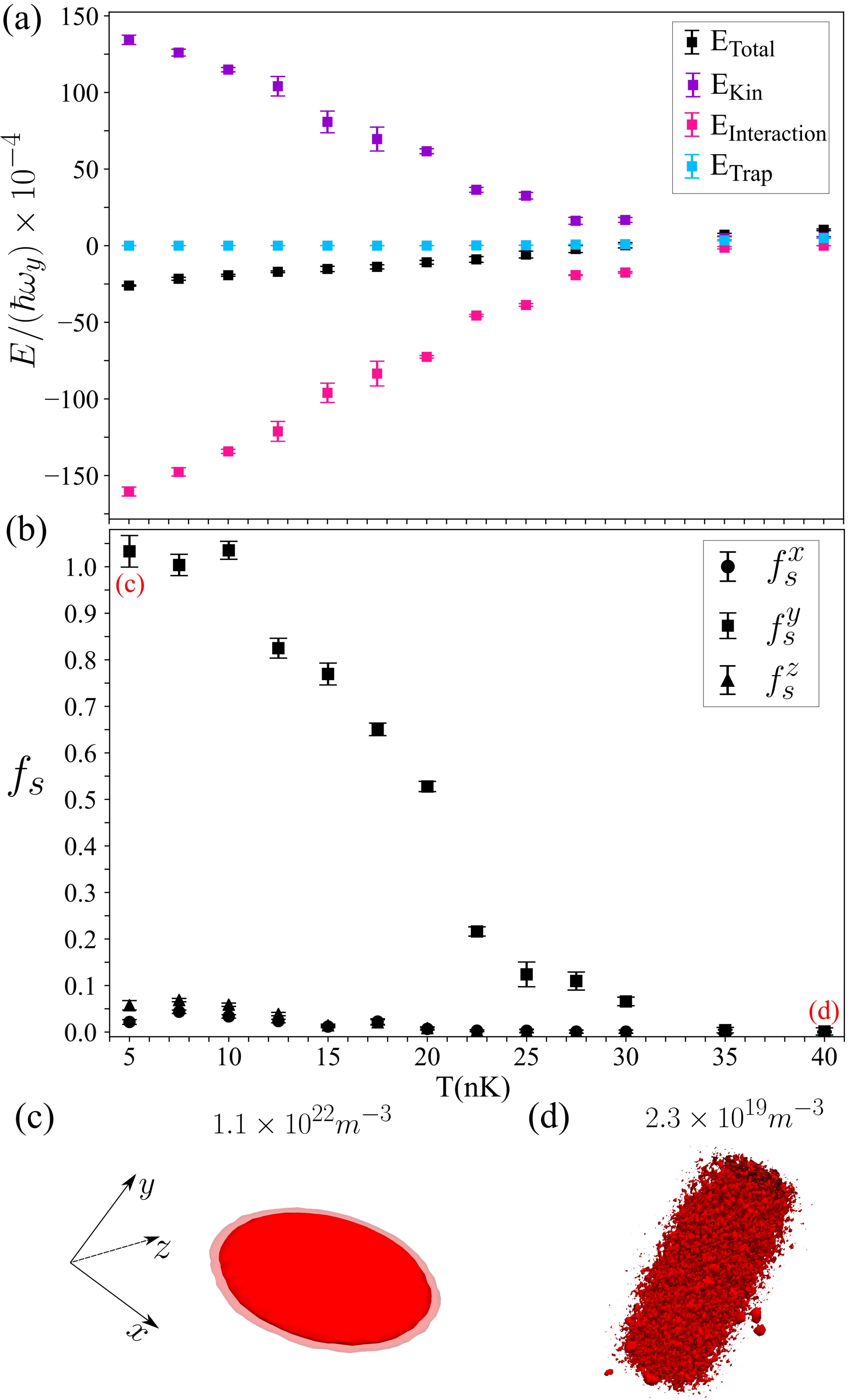}
    \caption{(Color online) (a)Energy variation with temperature for a pancake droplet at $a_s/a_0 = 170 $. The total energy is shown in black, the energy contributions arising from short-range interactions and long-range DDI are shown in pink, and the energy contributions from the kinetic term and external harmonic trap are illustrated in violet and skyblue respectively.  (b) Superfluid fraction along the transition from a thermal gas to a superfluid pancake droplet along three orthogonal directions($x,y,z$). The 3d isodensity surfaces are shown for (c) a superfluid pancake droplet (T = 5 nK), and (d) a cigar-shaped thermal gas (T = 40 nK). The peak densities for each case are provided at the top. The PIMC simulations are done for N = 1024 Dy atoms for an external harmonic trap, $\omega_{x,y,z}$ = $2\pi\times (75,25,75)$ Hz.} 
    \label{figure:5}
\end{figure}

 In the low scattering length regime, the doubly-dipolar gas is dominated by the strong DDI where the quantum fluctuations present in the system play a pivotal role. We investigate the interesting transition from a thermal gas to a quantum pancake droplet with varying temperatures.
 
 As expected, at high temperatures, thermal fluctuations are strong and we observe a cigar-shaped thermal gas (see Figure \ref{figure:5}d). As temperature decreases, the attractive DDI and the quantum fluctuations become prominent leading to the transition from the thermal gas to a quasi-2d pancake droplet. Figure \ref{figure:5}a shows the energy computed from PIMC simulations across this transition. The high quantum fluctuations at low temperatures are reflected in the increase of the kinetic energy of the system, while the total energy decreases due to diminished thermal fluctuations and an increase in the attractive component of the DDI. The effect of the external trap is minimal in this regime, which indicates the pancake droplet is effectively self-bound due to the DDI. 

The behavior of superfluidity across the temperatures is illustrated in Figure \ref{figure:5}b. The rise in the superfluid fraction around $T = 30$\, nK signals the formation of the quantum droplet. Upon entering the droplet regime, $f_s^{y}$ increases steadily and reaches $100\%$ superfluidity around $10$\,nK, while $f_s^{x,z}$ remains suppressed throughout due to the anisotropic DDI. The iso-density plot of a superfluid pancake droplet is shown in Figure~\ref{figure:5}c. Notably, we observe that the transition temperature from the thermal gas to the droplet is higher than that of the thermal gas to a condensate transition. This shows that the doubly-dipolar potential alters the critical temperature for the condensation in the Bose gas. With decreasing $a_s$ value, the DDI's dominance has shifted the critical temperature to a higher value. The pancake droplet demonstrates prolonged superfluidity across a broader temperature range compared to the condensate, which shows that the droplet is more robust against thermal fluctuations. 

Our study reveals that the doubly-dipolar condensate and the pancake droplet are stable against finite temperature fluctuations, which paves the way toward the experimental realization of DDBEC.

\section{Structural transition with Varying electric dipole moment}
\label{gamma}

\begin{figure}[t!]
    \centering
    \includegraphics[width=\columnwidth]{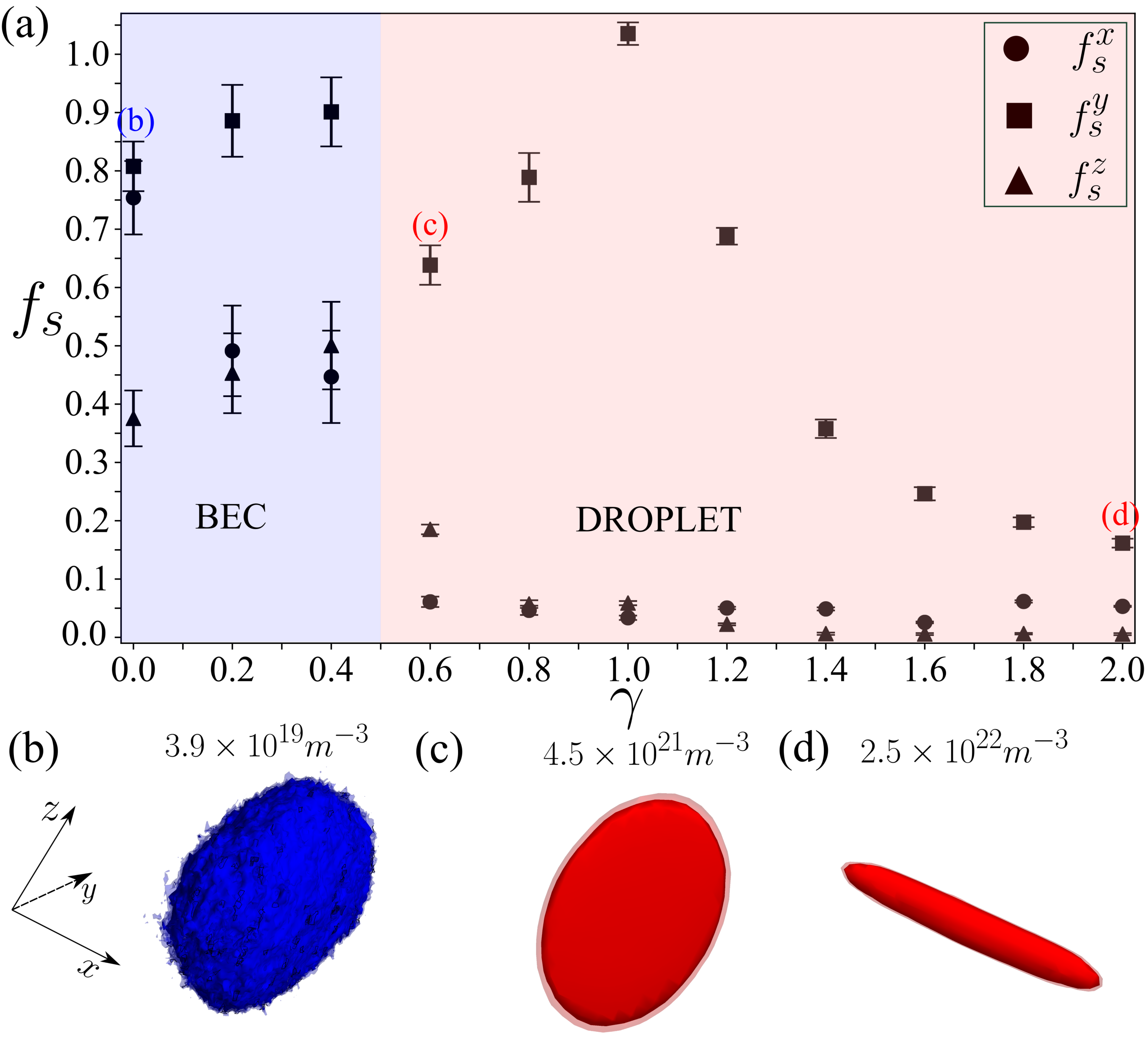}
    \caption{(Color online)(a) Superfluid fraction across the structural transition from a BEC to a quasi-2d pancake droplet to a cigar droplet along three orthogonal directions($x,y,z$) with increasing $\gamma$, for a constant $\alpha$ = $\pi/2$ at scattering length, $a_s = 170 a_0$. The 3d isodensity plots are shown for (b) BEC at $\gamma = 0$,(c) anisotropic pancake droplet at $\gamma = 0.6$  (d) cigar droplet elongated along $x$-axis at $\gamma = 2$. The peak densities for each case are provided at the top. The PIMC simulations are done for N = 1024 Dy atoms for an external harmonic trap, $\omega_{x,y,z}$ = $2\pi\times (75,25,75)$ Hz. The blue and red background denote the BEC and the droplet regime respectively.} 
    \label{figure:6}
\end{figure}

At this point, we study the characteristics of the doubly-dipolar Bose gas by manipulating the relative strength, $\gamma$, between the electric and magnetic dipole moments at a fixed scattering length.  $\gamma$ can be varied by varying the magnitude of the electric dipole moment with the help of the external electric field \cite{lepers2018}. The relative angle between the two dipole moments, $\alpha$, is kept constant at $\pi/2$.

The PIMC simulations with varying $\gamma$ reveal the BEC to droplet transition and the structural transition from a quasi-2d pancake to a quasi-1d cigar droplet. The superfluid behavior throughout this transition is illustrated in Figure \ref{figure:6}a. When $\gamma = 0 $, DDI solely involves magnetic dipoles pointing along the $z$-direction. The repulsive short-range interaction dominates over the attractive DDI and forms a repulsive BEC (see Figure \ref{figure:6}b). Since the trap is weaker along $y$, the condensate is elongated in $y$ and the width of the condensate along $z$ is greater than along $x$, due to the maximum attraction of DDI along $z$. The superfluid fraction exhibits finite values in all three orthogonal directions, with $f_s^{x,y}$  having higher values than $f_s^{z}$. With the introduction of $\gamma$ and the associated electric dipole moment, attractive interactions along the $x$-direction come into play. In the BEC regime for $\gamma> 0$, $f_s^{y}$ surpasses $f_s^{x,z}$ due to attractive DDI in the $xz$-plane. As $\gamma$ increases further, the attractive DDI dominates over the short-range interactions leading to the emergence of a quasi-2d pancake droplet. The pancake droplet is anisotropic for $\gamma \neq 1$, it is elongated along $z$ for $\gamma <1$ as shown in Figure \ref{figure:6}c. For $\gamma = 1$, where DDI is radially symmetric in the $xz$-plane, an isotropic pancake is formed, see Figure \ref{figure:5}c. Superfluidity is high perpendicular to the dipole plane for the pancake droplets and suppressed in the dipole plane. $f_s^{y}$ reaches around unity for an isotropic pancake droplet. For $\gamma$ values much higher than one, the electric dipole moment dominates over the magnetic dipole moment and DDI is maximally attractive along the $x$-axis. The droplet undergoes a structural transition from the pancake to a cigar shape elongated along the $x$-axis (see Figure \ref{figure:6}d). Superfluidity ($f_s^{y}$) decreases with the dimensional crossover from a quasi-2d to quasi-1d geometry. This study highlights that the superfluid characteristics of DDBEC can be manipulated through the application of an external field.
 
\section{Conclusions and outlooks} \label{con}

In this work, we have implemented first principles numerical simulations to study the properties of a doubly-dipolar Bose gas. Employing PIMC, we observe the transition from a repulsive BEC to a quantum pancake droplet at low temperatures and conduct a comparative study of the same at $T = 0$ using the gGPE in the mean-field framework. While PIMC and gGPE results align well in weak DDI regimes, PIMC correctly estimates dipolar as well as quantum fluctuation effects in a strong DDI regime. The superfluid fractions computed as a function of scattering lengths and temperatures reveal the highly anisotropic superfluid behavior of the pancake droplet. Notably, we observe that the temperature of passage from a thermal gas to the pancake droplet is higher than that of the transition to a BEC, showcasing the robustness of the pancake droplet against thermal fluctuations. Modulating the relative strength between the two dipole moments influences the structural and superfluid properties of the quantum droplet. Although our current focus has been on a single pancake droplet, by altering the trap geometry or number of atoms multiple pancake droplets can be realized. Future investigation of superfluidity between the droplets will provide insights into novel pancake supersolids predicted in the mean-field work \cite{rathee2022}. Exploring the finite temperature properties of pancake supersolids will also be the subject of future studies. Our study confirms a stable DDBEC at finite temperature and paves the way toward its experimental realization.

\begin{acknowledgements}
R. G. and R. N. thank the National Supercomputing Mission (NSM) for providing computing resources of “PARAM Brahma” at IISER Pune, which is implemented by C-DAC and supported by the Ministry of Electronics and Information Technology (MeitY) and Department of Science and Technology (DST), Government of India. This work was also supported by the European Union through the Next Generation EU funds through the Italian MUR National Recovery and Resilience Plan, Mission 4 Component 2 - Investment 1.4 - National Center for HPC, Big Data and Quantum Computing (CUP B83C22002830001).
M. C. and F. C. acknowledge financial support from PNRR MUR Project No. PE0000023-NQSTI. R.N. acknowledges DST-SERB for Swarnajayanti fellowship File No. SB/SJF/2020-21/19, and the MATRICS grant (MTR/2022/000454) from SERB, Government of India.

\end{acknowledgements}

\bibliography{libpimc.bib}

\appendix

\section{Density profile test for a single dipolar term}\label{densitytest}

\begin{figure}[h!]
    \centering
    \includegraphics[width=\columnwidth]{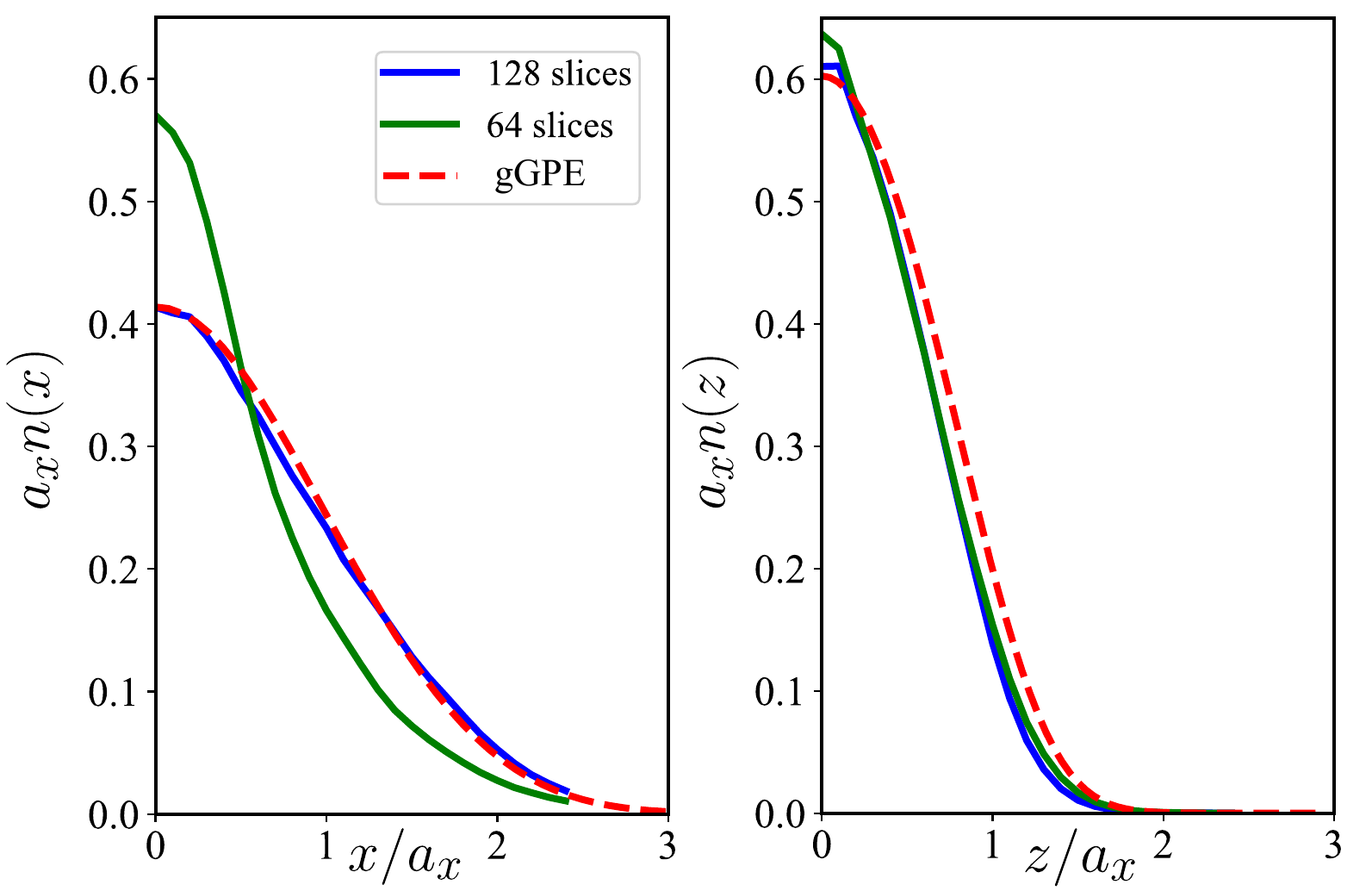}
    \caption{(Color online) Integrated density distribution along the $x$ and $z$-axis obtained using PIMC (solid lines) and gGPE (dashed lines) simulations. The simulations are done at $a_s=70 a_0$ for N = 1024 atoms under an external harmonic trap, $\omega_{x,y,z}$ = $2\pi\times (46,44,133)$ Hz.
    } 
    \label{fig:2}
\end{figure}

We have tested our PIMC code in the case of the simple dipolar interaction against GPE and known results. We consider $N=1024$ particles under an external harmonic confinement $\omega=2\pi(46,44,133)$Hz, with magnetic dipole moment $d_m=9.93\mu_b$. The s-wave scattering length, $a_s$ coincides with the radius $r_0$ of the hard-sphere potential. We run the PIMC code for a temperature $T=7.4$\,nK, taking $a_s=70 a_0$, where $a_0$ is the Bohr radius. Figure \ref{fig:2} depicts the integrated density profiles in the $x$ and $z$-directions for $M=64,128$ (see sec.~II). The PIMC density profiles are compared with the same quantity obtained by using the gGPE, which includes the beyond mean-field LHY correction. 
The density profiles of PIMC and gGPE find a good agreement with each other. At the same time, they too accord with the results by Saito for the same parameters \cite{saito2016}. 

\end{document}